\documentclass[prl,twocolumn,superscriptaddress]{revtex4}

\usepackage{bm}
\usepackage{graphics}
\usepackage{amsmath,epsfig}
\usepackage{amssymb}
\usepackage{comment}
\begin{document}



\title{Demonstrating anyonic fractional statistics with a six-qubit quantum simulator}

\author{Chao-Yang Lu}
\affiliation{Hefei National Laboratory for Physical Sciences at
Microscale and Department of Modern Physics, University of Science
and Technology of China, Hefei, 230026, China}
\author{Wei-Bo Gao}
\affiliation{Hefei National Laboratory for Physical Sciences at
Microscale and Department of Modern Physics, University of Science
and Technology of China, Hefei, 230026, China}
\author{Otfried G\"{u}hne}
\affiliation{Institut f\"{u}r Quantenoptik und Quanteninformation,
\"{O}sterreichische Akademie der Wissenschaften, Technikerstra{\ss}e
21A, A-6020 Innsbruck, Austria} \affiliation{Institut f\"{u}r
theoretische Physik, Universit\"{a}t Innsbruck, Technikerstra{\ss}e
25, A-6020 Innsbruck}
\author{Xiao-Qi Zhou}
\affiliation{Hefei National Laboratory for Physical Sciences at
Microscale and Department of Modern Physics, University of Science
and Technology of China, Hefei, 230026, China}
\author{Zeng-Bing Chen}
\affiliation{Hefei National Laboratory for Physical Sciences at
Microscale and Department of Modern Physics, University of Science
and Technology of China, Hefei, 230026, China}
\author{Jian-Wei Pan}
\affiliation{Hefei National Laboratory for Physical Sciences at
Microscale and Department of Modern Physics, University of Science
and Technology of China, Hefei, 230026, China}
\affiliation{Physikalisches Institut, Universit\"{a}t Heidelberg,
Philosophenweg 12, 69120 Heidelberg, Germany}

\date{\today}

\begin{abstract}
Anyons are exotic quasiparticles living in two dimensions that do
not fit into the usual categories of fermions and bosons, but obey a
new form of fractional statistics. Following a recent proposal
[Phys. Rev. Lett. \textbf{98}, 150404 (2007)], we present an
experimental demonstration of the fractional statistics of anyons in
the Kitaev spin lattice model using a photonic quantum simulator. We
dynamically create the ground state and excited states (which are
six-qubit graph states) of the Kitaev model Hamiltonian, and
implement the anyonic braiding and fusion operations by single-qubit
rotations. A phase shift of $\pi$ related to the anyon braiding is
observed, confirming the prediction of the fractional statistics of
Abelian $1/2$-anyons.
\end{abstract}
\pacs{Valid PACS appear here}

\maketitle

Quantum statistics classifies fundamental particles in three
dimensions as bosons and fermions. Interestingly, in two dimensions
the laws of physics permit the existence of exotic
quasiparticles---anyons---which obey a new statistical behavior,
called fractional or braiding statistics \cite{77}. That is, upon
exchange of two such particles, the system wave function will
acquire a statistical phase which can take any value---hence this
name. Anyons have been predicted to live as excitations in
fractional quantum Hall (FQH) systems \cite{tsui,laughlin,wen}.
Alternatively, quantum states with anyonic excitations can be
artificially designed in spin model systems that possess highly
nontrivial ground states with topological order. A prominent example
is the Kitaev spin lattice model \cite{kitaev1,kitaev2}, which
opened the avenue of fault-tolerant topological quantum computing
\cite{franktoday,why}.

It is an important goal to manipulate the anyons and demonstrate
their exotic statistics. Towards this goal, a number of theoretical
schemes have been proposed, both in the FQH regime \cite{hall} and
in the Kitaev models \cite{duan,zoller,zhang,jiang,han}. However, it
has proved extremely difficult to experimentally detect the
fractional statistics associated with anyon braiding. While recent
experiments in the FQH systems have indeed revealed some signatures
of anyonic statistics \cite{prb,signiture}, resolving individual
anyons remains elusive \cite{franktoday}. Here we take a different
approach to this challenge --- exploiting the spin models to study
the anyonic statistics. Following a recent proposal \cite{han}, we
demonstrate the fractional statistics of anyons by simulation of the
Kitaev model on a six-photon graph state. The method is to
dynamically create the ground state and excited state of the anyonic
model Hamiltonian, and implement the braiding and fusion operations
by single-qubit rotations.

How can the statistical nature of elementary particles be
experimental observable? First let us recall the concept of quantum
statistics. The wave function of a two-particle system
$\psi(\mathbf{r}_1,\mathbf{r}_2)$ will acquire a statistical phase
$\theta$ upon an adiabatic exchange of two particles, that is,
$\psi(\mathbf{r}_2,\mathbf{r}_1)=e^{i\theta}\psi(\mathbf{r}_1,\mathbf{r}_2)$,
where $\theta=0$ for bosons, $\theta=\pi$ for fermions, and $\theta$
can be any value ($0\leqslant\theta\leqslant\pi$) for anyons. It can
be seen that a full circulation of a particle around the other one
is equivalent to two successive particle exchanges \cite{why}. After
such a circulation, both bosons and fermions show a trivial phase
($\phi=2\theta=0,2\pi$), but anyons will get an observable
non-trivial phase $\phi$. To realize this idea, we need a system
where anyons can be created and braiding operations can be carried
out experimentally. The Kitaev model is well suited for this.

The first Kitaev model was designed on a spin lattice with qubits
living on the edges (see Fig. 1a). For each vertex $v$ and face $f$,
we consider operators of the form
\begin{equation}\label{1}
A_v\,=\prod_{j\in \mathrm{star}(v)}X_j \, , \quad B_f\,=\prod_{j\in
\mathrm{boundary}(f)}Z_j,
\end{equation}
where $X$ ($Z$) denotes the standard Pauli matrix $\sigma_x$
($\sigma_z$). These operators $A_v$, $B_f$ are put together to make
up the model Hamiltonian
\begin{equation}\label{2}
H\,=\,-\sum_{v}A_v-\sum_{f}B_f.
\end{equation}
The ground state $|\psi_g\rangle$ of this Hamiltonian (\ref{2}) is
given by $A_v|\psi_g\rangle=|\psi_g\rangle$ and
$B_f|\psi_g\rangle=|\psi_g\rangle$ for all vertices and faces.
Violations of these conditions cost energy and generate excited
states $|\psi_{e}\rangle$. A quasiparticle is created on the vertex
$v_i$ (face $f_i$) if $A_{v_i}$ ($B_{f_i}$) acting on the excited
state $|\psi_{e}\rangle$, yields an eigenvalue $-1$ instead of $+1$
for the ground state. In Ref. \cite{kitaev1,han}, the quasiparticles
on vertices are called ``electric charges'' (e-particles for short)
and those on faces are called ``magnetic vortices'' (m-particles).
It is shown that these particles have unusual mutual statistical
properties, as we can get a phase flip $-1$ if we move one particle
around the other, which are thus called abelian $1/2$-anyons
\cite{kitaev1} (see Fig. 1b).

\begin{figure}[bt]
\centering
  \includegraphics[width=0.46\textwidth]{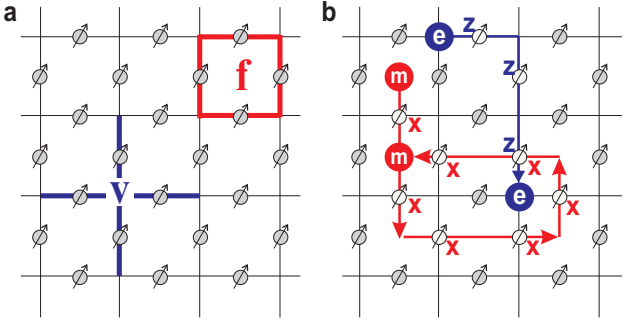}\\
  \caption{The first Kitaev model \cite{kitaev1} and anyonic braiding operations \cite{han}.
  (\textbf{a}). In this spin lattice, the qubits live on the
  edges and the stabilizer operators $A_v$, $B_f$
  (\ref{1}) represent the four-body interactions as illustrated.
  (\textbf{b}). Creation of quasiparticles and braid of an m-particle around an
  e-particle. A pair of e-particles (m-particles) are created on vertices (faces) by
  applying a $Z$ ($X$) operation on the edge qubit. The
  quasiparticles can be moved horizontally and vertically by
  repeated applications of $Z$ or $X$ operations. The figure shows
  an example of how
  an m-particle forms a closed loop around an e-particle through a series of
  moves.
     }\label{}
\end{figure}

Recently, Han, Raussendorf and Duan \cite{han} exploited the fact
that the statistical properties of anyons are manifested by the
underlying ground and excited states \cite{levin}. So, instead of
direct engineering the interactions of the Hamiltonian $H$ (\ref{2})
and ground-state cooling which are extremely demanding
experimentally, an easier way is to dynamically create the ground
state and the excitations of this model Hamiltonian, encoding the
underlying anyonic model in a multiparticle entangled state which
can used to simulate the dynamics of the anyonic system. The
quasiparticles are then defined by the negative outcome of a
stabilizer element $A_v$ or $B_f$. Specificaly, as illustrated in
Fig. 1b, with the ground state $|\psi_g\rangle$ prepared, one can
first create a pair of e-particles by applying a single-qubit $Z$
rotation. The system wave function will be in the excited state
$|\psi_e\rangle$. To make fractional phase experimentally detectable
in a later stage, we apply a $\sqrt{Z}$ rotation and get a
superposition state $(1/\sqrt{2})(|\psi_g\rangle+|\psi_e\rangle)$.
Then we create another pair of m-particles and move one of them
around an e-particle along a closed loop, and finally annihilate the
m-particles. After doing so, it is predicted that a fractional phase
$\pi$ will be added to $|\psi_e\rangle$, thus the superposition
state will become $(1/\sqrt{2})(|\psi_g\rangle-|\psi_e\rangle)$.

As the anyons are perfectly localized quasiparticles in this model
Hamiltonian, a small spin lattice containing six qubits shown in
Fig. 2a allows for a proof-of-principle experimental demonstration
\cite{han,kitaev1}. The Hamiltonian of this model is
$H_6=-A_1-A_2-B_1-B_2-B_3-B_4$, where $A_1=X_1X_2X_3$,
$A_2=X_3X_4X_5X_6$, $B_1=Z_1Z_3Z_4$, $B_2=Z_2Z_3Z_5$, $B_3=Z_4Z_6$,
$B_4=Z_5Z_6$ (the subscripts of the Pauli matrices label the
qubits). The ground state $|\psi\rangle_6$ of this Hamiltonian
$H_6$ is locally equivalent to a six-qubit graph state 
\cite{cluster,graph}, which can be represented by the graph as
depicted in Fig. 2b. This equivalence follows from the fact that the
operators $A_1$, $\cdots$, $B_4$ in the Kitaev model can be uniquely
derived from the stabilizer operators $g_i$ (see Fig. 2b) of the
graph state.

\begin{figure}[tb]
\centering
  \includegraphics[width=0.49\textwidth]{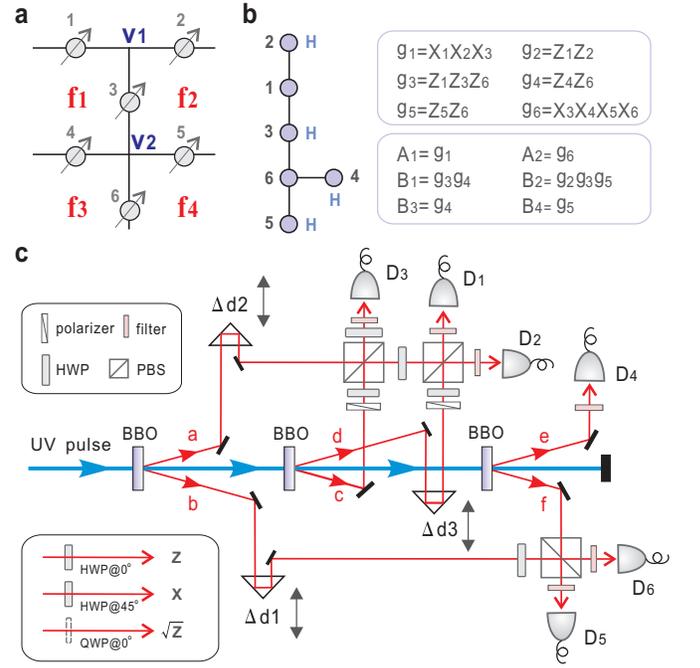}\\
  \caption{
  (\textbf{a}). The small Kitaev spin lattice system with six qubits used for demonstration of anyonic
  braiding operations.
  (\textbf{b}). The six-qubit graph state which, after Hadamard (H) transformations on
  qubit $2$, $3$, $4$, and $5$,
  is equivalent to the ground state of 
  the system in Fig. 2a.
  The graph state is associated with a graph, where each vertex
  denotes a qubit prepared in the state $(1/\sqrt{2})(|0\rangle+|1\rangle)$ and
  each edge represents a controlled phase gate having been applied
  between the two connected qubits \cite{cluster,graph}. The graph state is a common
  eigenstate of the stabilizer operators $g_i$, that is,
  $g_i|\psi\rangle_6=|\psi\rangle_6$, which describe the correlation in the state, and the graph state is the unique
  state fulfilling this. Here we use the same label as Fig. 2a-b in ref. \cite{han}. (\textbf{c}). Experimental set-up for the
  generation of graph state and demonstration of braiding
  operations. A pulsed ultraviolet laser successively passes
  through three $\beta$-barium borate (BBO) crystals to generate
  three pairs of entangled photons \cite{kwiat}.
  The photons $a$, $b$, $c$, $d$, and $f$ are combined on the three
  polarizing beam splitters (PBSs) step by step \cite{pan2001}. To achieve good spatial and temporal overlap,
  all photons are spectrally filtered
  ($\Delta\lambda_{\mathrm{FWHW}}=3.2\mathrm{nm}$) and detected by
  fiber-coupled single-photon detectors ($\mathrm{D}_1, \cdots, \mathrm{D}_6$). 
   The detector labels
correspond to the qubit labels in the text and in Fig. 2a-b. The
coincidence events are registered by a
  programmable multichannel coincidence unit. For single-qubit rotations and polarization
  analysis, quarter wave plates (QWPs), half wave plates (HWPs),
  together with polarizers or PBSs are used.
     }\label{}
\end{figure}

Now we proceed with the experiment in three steps: (1) create and
analyze of the ground state $|\psi\rangle_6$, (2) verify the anyonic
excitations, (3) implement the braiding operations and  detect the
anyonic phase. We use single photons as a real physical system to
simulate the creation and control the anyons. The quantum states are
encoded in the polarization of the photons which are robust to
decoherence. The experimental set-up is illustrated in Fig. 3c. We
start from three pairs of entangled photons produced by spontaneous
parametric down-conversion (SPDC) \cite{kwiat}. The photons in
spatial modes $a$-$b$ and $e$-$f$ are prepared in the states
$|\phi^+\rangle_{ij}=(1/\sqrt{2})(|H\rangle_i|H\rangle_j+|V\rangle_i|V\rangle_j)$,
while those in mode $c$-$d$ are disentangled using polarizers and
then prepared in the states
$|+\rangle_i=(1/\sqrt{2})(|H\rangle_i+|V\rangle_i)$, where $H$($V$)
denotes horizontal (vertical) polarization, and $i$ and $j$ label
the spatial modes. We then pass the photons through a linear optics
network (see Fig. 3c). A coincidence detection of all six outputs
corresponds exactly to the ground state
\begin{eqnarray}
   \nonumber |\psi\rangle_6&=&\frac{1}{2}\,(|H\rangle_1|H\rangle_2|H\rangle_3|H\rangle_4|H\rangle_5|H\rangle_6  \\
   \nonumber &&+\,\,|V\rangle_1|V\rangle_2|V\rangle_3|H\rangle_4|H\rangle_5|H\rangle_6    \\
   \nonumber &&+\,\,|H\rangle_1|H\rangle_2|V\rangle_3|V\rangle_4|V\rangle_5|V\rangle_6  \\
    &&+\,\,|V\rangle_1|V\rangle_2|H\rangle_3|V\rangle_4|V\rangle_5|V\rangle_6).
 \end{eqnarray}

To verify that the ground state $|\psi\rangle_6$ has been obtained,
first we experimentally measure the expectation values of its
stabilizer operators $A_1$, $\cdots$, $B_4$. These stabilizer
operators describe the intrinsic correlations in the state
$|\psi\rangle_6$ and uniquely define it, thus their expectation
values could serve as a good experimental signature. For an ideal
state $|\psi\rangle_6$, all expectation values should give $+1$. In
our experiment however, the ground state was created imperfectly.
Figure 3a shows the measurement results, with all expectation values
being positive in a rang between $0.51\pm0.04$ and $0.74\pm0.03$, in
qualitative agreement with the theoretical prediction. For a more
complete and quantitative analysis, we aim to estimate the fidelity
of the produced state, that is, its overlap with the desired one.
This quantity is given by
$F_{\psi_6}={_6}\langle\psi|\rho_{\mathrm{exp}}|\psi\rangle_6$,
which is equal to one for an ideal state, and $1/64$ for a
completely mixed six-qubit state. To do so, we consider a special
observable, which allows for lower bounds on the fidelity, while
being easily measurable with few correlation measurements
\cite{otfried, epaps}. By making these measurements, we estimate the
fidelity of the created ground state to be
$F_{\psi_6}\geqslant0.532\pm0.041$ \cite{note}. The imperfection of
this state is mainly caused by the high-order photon emissions
during the SPDC and the partial distinguishability of independent
photons \cite{white}.

We now move to the step (2). With the ground state $|\psi\rangle_6$
created, we apply a $Z$ (X) rotation on qubit $3$ ($4$), creating an
excited state $|\psi_{em}\rangle_6$ on which a pair of e-particles
live on the vertices $v_1$ and $v_2$, and another pair of
m-particles on faces $f_1$ and $f_3$ (see Fig. 2a). The $Z$ and $X$
rotations are experimentally realized using HWPs oriented at
$0^{\circ}$ and $45^{\circ}$, respectively. As discussed before, the
anyonic excitations are signaled by violations of stabilizer
conditions, that is,
$A_{v_i}|\psi_{em}\rangle_6=-|\psi_{em}\rangle_6$,
$B_{f_i}|\psi_{em}\rangle_6=-|\psi_{em}\rangle_6$ \cite{kitaev1}.
Thus in our case, theoretically, the expectation values of $A_1$ and
$A_2$ should become $-1$ because of the e-particles, and the same
for $B_1$ and $B_3$ due to the m-particles. To verify this, we
measure the expectation values of the operators
$A_1$,$\cdots$,$B_4$. The results are shown in Fig. 3b, where the
values of $A_1$, $A_2$, $B_1$ and $B_3$ flip compared to those shown
in Fig. 3a which supports the presence of anyonic excitations
\cite{kitaev1,han}.

\begin{figure}[tb]
\centering
  \includegraphics[width=0.48\textwidth]{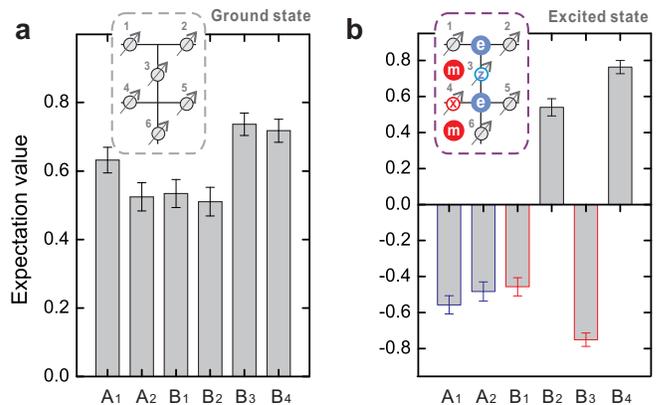}\\
  \caption{The measured expectation values of the operators $A_1$,
  $\cdots$, $B_4$ of the ground state $|\psi\rangle_6$ (\textbf{a})
and the excited state $|\psi_{em}\rangle_6$ (\textbf{b}). The
excited state $|\psi_{em}\rangle_6$ has a pair of e-particles on
$v_1$, $v_2$ and a pair of m-particles on $f_1$, $f_3$, thus the
values for $A_1$, $A_2$, $B_1$, $B_3$ become negative. Each
expectation value is derived from a complete set of $64$ six-fold
coincidence events in $15$$\mathrm{h}$ in measurement basis
$Z^{\otimes 6}$ or $X^{\otimes 6}$. The error bars represent one
standard deviation, deduced from propagated Poissonian statistics of
the raw detection events.
     }\label{}
\end{figure}

Now we proceed to the step (3). On the ground state
$|\psi\rangle_6$, first we apply a $\sqrt{Z}$ operation using a QWP
oriented at $0^{\circ}$ on the qubit $3$ of the ground state
$|\psi\rangle_6$, yielding a superposition state
$|\psi_s\rangle_6=(1/\sqrt{2})(|\psi\rangle_6+|\psi_e\rangle_6)$,
where $|\psi_e\rangle_6$ is the excited state with a pair of
e-particles on $v_1$ and $v_2$. With an $X$ rotation on the qubit
$4$ we further create a pair of m-particles on $f_1$ and $f_3$. Then
we perform four $X$ operations on the qubits $6$-$5$-$3$-$4$ to
implement the braiding operation, that is, the m-particle on $f_3$
is moved around the e-particle on $v_2$ along an anticlockwise
closed loop. We note that the crossing at the qubit $3$, which is
unavoidable in two dimensions, is relevant for the unusual
statistics. Finally, the pair of m-particles is annihilated with an
$X$ operation on qubit $4$ (fusion).

After these, if there is a fractional phase $\phi$ acquired, the
state $|\psi_s\rangle_6$ will become
$|\psi_f\rangle_6=(1/\sqrt{2})(|\psi\rangle_6+e^{i\phi}|\psi_e\rangle_6)$.
To determine this $\phi$, we look at the correlation measurement
outcomes of the six photons where the photons $1$ and $2$ are fixed
at $|+\rangle$ polarization, $4$, $5$ and $6$ at $|H\rangle$ and the
photon $3$ is measured in the basis
($|+\rangle+e^{i\alpha}|-\rangle$) with $\alpha$ varying in $\pi/4$
steps. In this setting, the six-fold coincidence counts should
follow the relation $C(\phi,\alpha)\propto1+\sin(\phi-\alpha)$ for
the state $|\psi_f\rangle_6$, thus an unknown phase $\phi$, if
occurs, can be revealed. Figure 4a shows the measurement results for
both the state $|\psi_s\rangle_6$ and $|\psi_f\rangle_6$, before and
after the process of m-particle creation, braiding and fusion. These
two curves clearly exhibit a phase difference of $\pi$, confirming
the prediction of the fractional statistics.

For a more complete proof, we implement a $\sqrt{Z}$ transformation
on the qubit $3$ of the remaining state $|\psi_f\rangle_6$. The
state $|\psi_f\rangle_6$ will be converted to $|\psi\rangle_6$ if
there is a fractional phase $\pi$, otherwise it will go to
$|\psi_e\rangle_6$. To test this, again we measure the expectation
values of the operators $A_1$,$\cdots$,$B_4$ of the state after the
$\sqrt{Z_3}$ transformation. The experimental results are shown in
Fig. 4b, which are in agreement with that the final state is
$|\psi\rangle_6$ and thus prove the fractional phase change of
$\phi=\pi$. Here we note that the facts that the present setup use
free-flying non-interacting photons and that the timescale of the
braiding operations is extremely small ($\sim$ picoseconds for
photons passing through the HWPs and QWPs) implies that this
acquired phase cannot be a dynamical phase. Moreover, the creating,
braiding and annihilating of the m-particles which corresponds to
the operation $X_4X_6X_5X_3$ do not give rise to a phase either, as
the ground state $|\psi\rangle_6$ is an eigenstate of this
observable $X_4X_6X_5X_3$. Similar arguments also apply to the
e-particles' case. Consequently, this excludes possible geometrical
phases \cite{phase} and proves the observed phase is purely
statistical.

\begin{figure}[tb]
\centering
  \includegraphics[width=0.48\textwidth]{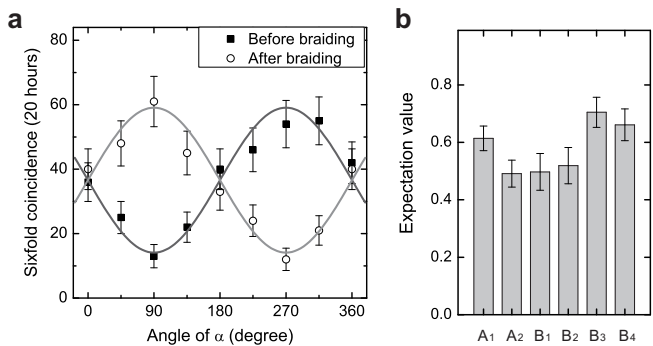}\\
  \caption{(\textbf{a}). Measured fringes for the state $|\psi_s\rangle_6$ and
  $|\psi_f\rangle_6$. The measurements in basis
  ($|+\rangle+e^{i\alpha}|-\rangle$) are done using a combination
  of HWPs, QWP and PBS. (\textbf{b}). The expectation values of the operators $A_1$,
  $\cdots$, $B_4$ of the state $|\psi_f\rangle_6$ after the
  $\sqrt{Z_3}$ transformation.
     }\label{}
\end{figure}

In summary, we have demonstrated the creation and manipulation of
anyons in the Kitaev spin lattice model, and observed the fractional
statistics of the abelian $1/2$-anyons. This has been done without
generating the four-body interactions in the model Hamiltonian but
alternatively, in an easier way---by encoding the underlying anyonic
system on a six-photon graph state, or equivalently, realizing the
six-qubit circuit shown in Fig. 2c of ref. \cite{han}. It should be
noted that the absence of the Hamiltonian does not prevent us from
studying the topological and statistical properties of the anyons
here; however, topological quantum computing in a fault-tolerant
manner would eventually require such a Hamiltonian. From a
quantum-information prospective, our experiment may be seen as using
quantum computers which have already well understood physics as a
tool to simulate other difficult quantum systems. Such quantum
simulators can in principle provide exponential speedup in the
simulation of quantum physics \cite{science}, and may offer a more
controllable and clean access to study strongly correlated behaviors
than natural complex solid-state systems. Possible future work along
this line may include investigation of the robustness of anyonic
braiding \cite{han} and other possible phases \cite{phase} using
hyper-entangled graph states \cite{ten}, and demonstration of some
basic elements of cluster-state topological quantum computing
\cite{cluster-topo}.

Acknowledgements: We thank R. Raussendorf and L.-M. Duan for helpful
discussions. This work was supported by the NNSFC, the CAS, the NFRP
(under Grant No: 2006CB921900), the A. v. Humboldt Foundation and
Marie Curie Excellence Grant of the EU, the FWF, the EU (Scala,
Olaqui, QICS).


\end{document}